\documentclass{camera}
\usepackage{graphicx}  % uncomment this if your want to insert figures
\usepackage[numbers, sort&compress]{natbib}
\begin{document}

%%%%%%%%%%%%%%%%%%%%%%%%%%%%%%%%%%%%%%%%%%%%%%%%%%%%%%%%
% The title, only the first letter capitalized; if you want to split it in
% two or more lines, put a \\ macro at each line break
% example: 
%   \title{Title: first line\\ second line}
%
\title{Quasifission dynamics in TDHF}

%%%%%%%%%%%%%%%%%%%%%%%%%%%%%%%%%%%%%%%%%%%%%%%%%%%%%%%%
% The author(s), separated by commas; do not put a
% comma before the last author, use instead the \and
% macro which produces a normal ``and'' in the
% caps/small caps context
%
\author{A.S. Umar$^1$, V.E. Oberacker$^1$ \and C. Simenel$^2$}

%%%%%%%%%%%%%%%%%%%%%%%%%%%%%%%%%%%%%%%%%%%%%%%%%%%%%%%%
%
\organization{$^1$Department of Physics and Astronomy, Vanderbilt University, Nashville, Tennessee 37235, USA\\$^2$Department of Nuclear Physics, RSPE, Australian National University, Canberra, ACT 0200, Australia}

\maketitle

\begin{abstract}
For light and medium mass systems the capture cross-section may be considered
to be the same as that for complete fusion, whereas for heavy systems leading to superheavy
formations the evaporation residue cross-section is dramatically reduced due to the
quasifission (QF) and fusion-fission processes thus making the capture cross-section
to be essentially the sum of these two cross-sections, with QF occurring at a much shorter time-scale. Consequently, quasifission is the primary reaction 
mechanism that limits the formation of superheavy nuclei.
Within the last few years the time-dependent Hartree-Fock (TDHF) approach has been utilized for studying the dynamics of quasifission.
The study of quasifission is showing a great promise to provide
insight based on very favorable comparisons with experimental data. 
In this article we will
focus on the TDHF calculations of quasifission observables for the
$^{48}$Ca+$^{249}$Bk system.
\end{abstract}

%%%%%%%%%%%%%%%%%%%%%%%%%%%%%%%%%%%%%%%%%%%%%%%%%%%%%%%%
% Write the text starting from here and using the usual
% LaTeX commands.
%
\section{Introduction}

One of the most fascinating research areas involving low-energy nuclear reactions
is the search for superheavy elements.
Experimentally, two approaches have been used for the synthesis of these elements,
one utilizing %closed shell nuclei with lead-based 
targets in the lead region (cold-fusion)~\cite{hofmann2000,hofmann2002},
the other utilizing deformed actinide targets with $^{48}$Ca projectiles 
(hot-fusion)~\cite{oganessian2007,oganessian2013,hofmann2007}.
While both methods have been successful in synthesizing new elements the evaporation
residue cross-sections of the hot-fusion reactions were found to be as much as three
times larger than those of the cold fusion ones.
To pinpoint the root of this difference it is important to understand the details
of the entrance channel dynamics of these systems since the properties of the
dinuclear system at the capture point will strongly influence the outcome of the
reaction.
For light and medium mass systems the capture cross-section may be considered
to be the same as that for complete fusion, whereas for heavy systems leading to superheavy
formations the evaporation residue cross-section is dramatically reduced due to the
quasi-fission \cite{Toke1985} and fusion-fission processes, thus making the capture cross-section to
be essentially the sum of these two cross-sections.
What is also difficult to ascertain is the configuration of the composite system, 
namely, whether the system has a single-center compound-like configuration
or a dinuclear configuration accompanied by particle exchange.
Most dynamical models~\cite{fazio2005,adamian2003,adamian2009,nasirov2009a,feng2009} argue that for heavy systems a dinuclear
complex is formed initially and  the barrier structure and the excitation energy of this precompound
system will determine its survival to breaking up via  quasi-fission.
Furthermore, if the nucleus survives this initial state and evolves to a compound system
it can still fission due to its excitation.

Within the last few years the time-dependent Hartree-Fock (TDHF) approach~\cite{negele1982,simenel2012} has been 
utilized for studying the dynamics of
quasifission~\cite{golabek2009,kedziora2010,wakhle2014,oberacker2014,kalee2015}
and scission dynamics~\cite{simenel2014a,scamps2015,goddard2015}. Particularly, the study of quasifission is showing a great promise to provide
insight based on very favorable comparisons with experimental data. 
Similarly, an extension of TDHF called the density-constrained TDHF~\cite{umar2006b}
(DC-TDHF) has
been used to obtain microscopic potential barriers and capture cross-sections for
superheavy~\cite{umar2010a} and lighter systems~\cite{oberacker2010,keser2012,simenel2013a}.
In this article we will
focus on the TDHF studies of quasifission for the $^{48}$Ca+$^{249}$Bk system.

\section{Results}
During the past several years it has become feasible to perform TDHF calculations
on a three-dimensional (3D) Cartesian grid with no symmetry restrictions
and with much more accurate numerical methods~\cite{bottcher1989,maruhn2014}.
In the present TDHF calculations we use the Skyrme SLy4d
energy density functional (EDF)~\cite{kim1997}
including all of the relevant time-odd terms in the mean-field Hamiltonian.
First we generate very accurate static HF wave functions for the two nuclei on the
3D grid.
The initial separation of the two nuclei is $30$~fm. In the second
step, we apply a boost operator to the single-particle wave functions. The time-propagation
is carried out using a Taylor series expansion (up to orders $10-12$) of the
unitary mean-field propagator, with a time step $\Delta t = 0.4$~fm/c.
By virtue of long contact-times for quasifission and the energy
and impact parameter dependence these calculations require extremely long CPU times.
\begin{figure}[!htb]
	\centerline{
	\includegraphics*[scale=0.3]{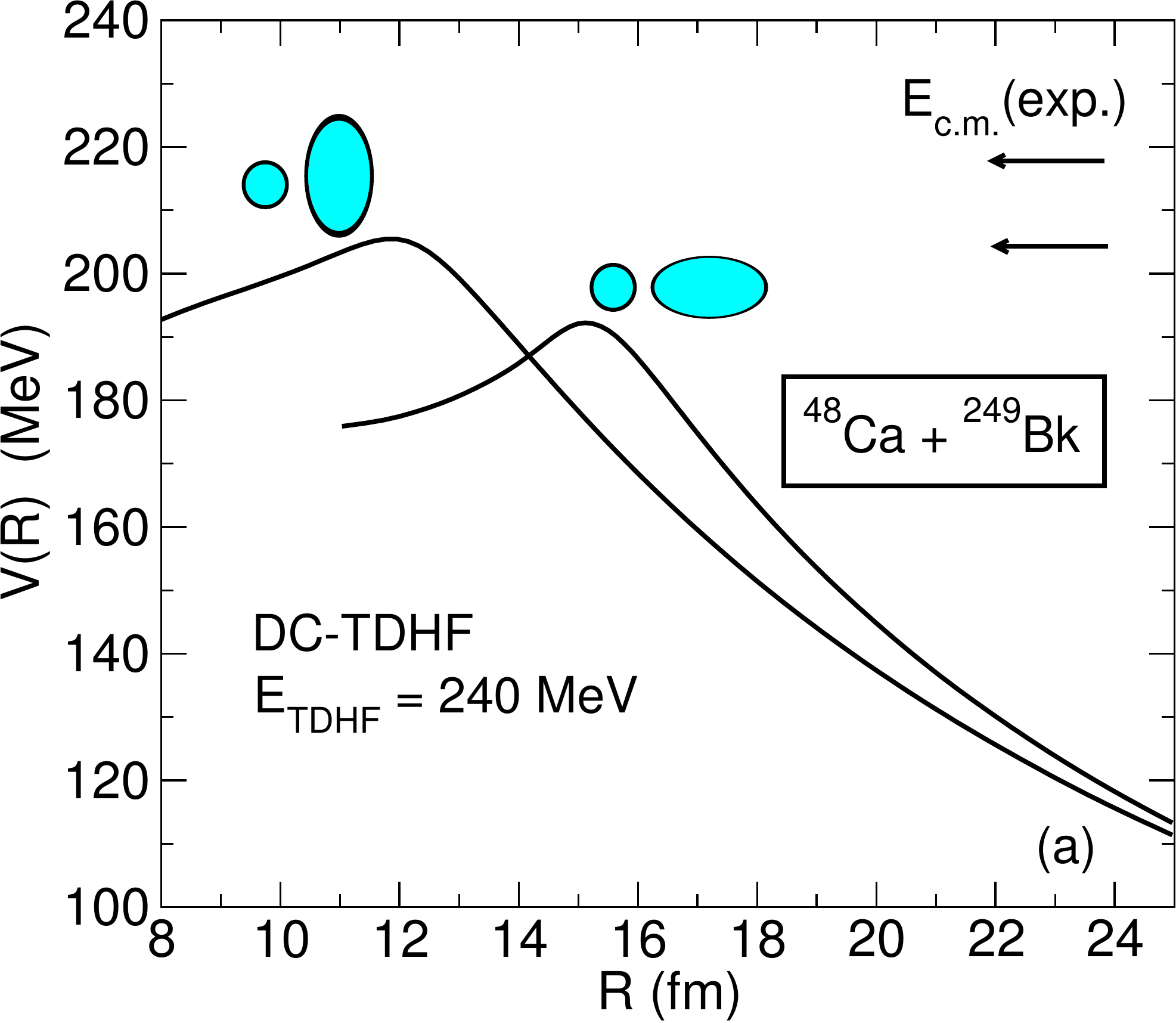}\hspace{0.2in}
	\includegraphics*[scale=0.3]{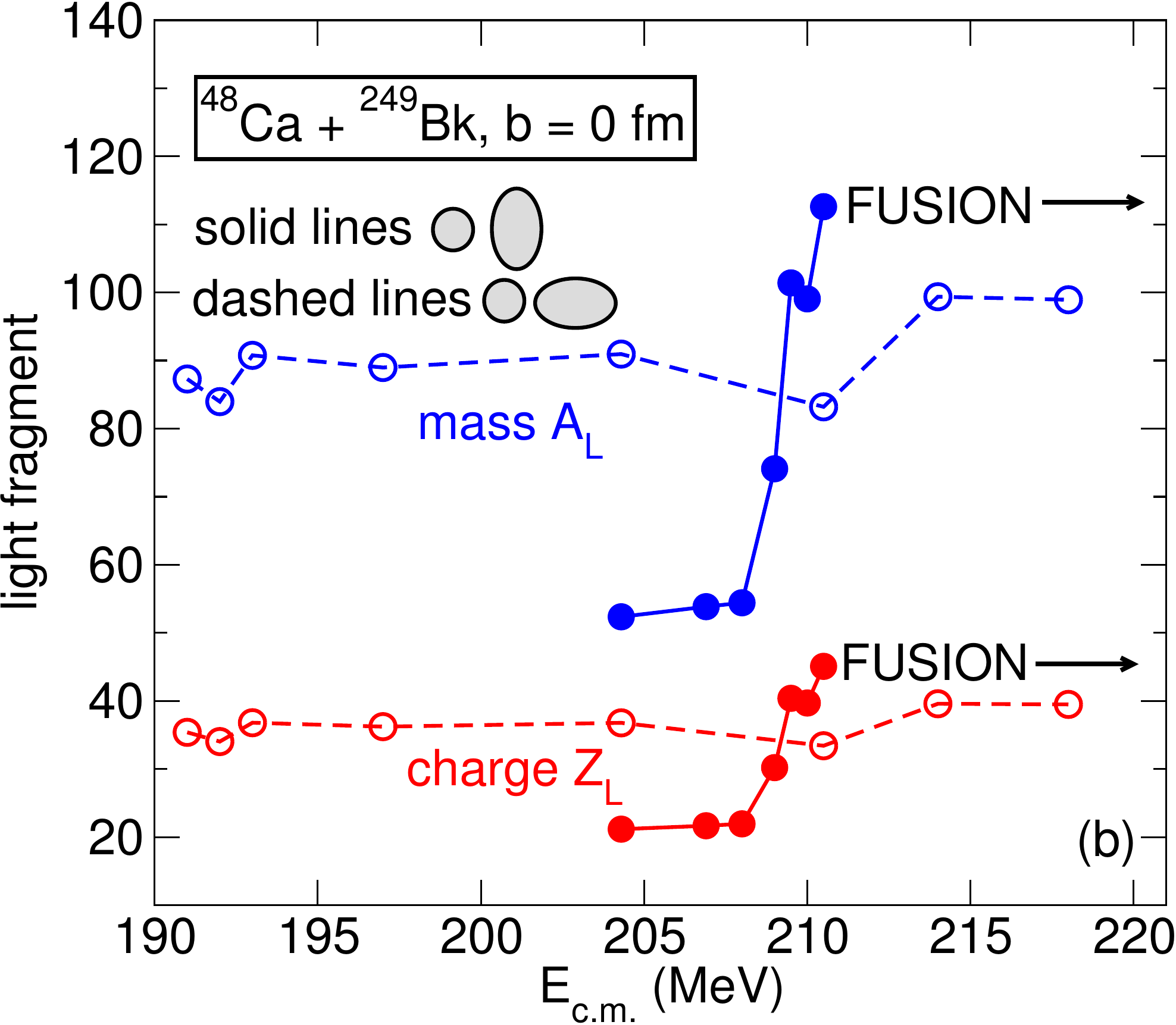}}
	\caption{\protect (a) Nucleus-nucleus potential, $V(R)$,  for the $^{48}$Ca+$^{249}$Bk system
		obtained from DC-TDHF calculation for selected orientation angles of the $^{249}$Bk nucleus. Also shown are the experimental c.m. energies.
		(b) Mass and charge of the light fragment as a function of $E_\mathrm{c.m.}$
		for central collisions of $^{48}$Ca with the side and tip orientations of $^{249}$Bk.}
	\label{fig:fig0}
\end{figure}
In Fig.~\ref{fig:fig0}a we plot the microscopic DC-TDHF potential barriers obtained
for the $^{48}$Ca+$^{249}$Bk system. The two barriers depict the two extreme orientations
of the $^{249}$Bk nucleus. Also, shown are the experimental energies at which this reaction has been studied~\cite{oganessian2013,khuyagbaatar2014}.
As expected the polar or tip orientation of $^{249}$Bk results in a significantly lower
barrier. The highest experimental energy is
above both barriers but the lowest experimental energy is below the barrier for the
equatorial or side orientation of $^{249}$Bk.
Figure~\ref{fig:fig0}b shows the mass and charge of the light fragment as a function of $E_\mathrm{c.m.}$
for central collisions of $^{48}$Ca with the side and tip orientation of $^{249}$Bk. For the side orientation of $^{249}$Bk for energies below
$E_\mathrm{c.m.}=204$~MeV we get quasielastic collisions whereas for energies above 211~MeV we see
fusion, which we define arbitrarily as reactions with contact times exceeding 35~zs. Naturally, non-central impact
parameters can show quasifission in the range where we see fusion.
We observe that for central collisions with the side orientation of $^{249}$Bk
quasifission is limited to a small range of energies $E_\mathrm{c.m.}=209-211$~MeV.
The quasifission results are very different for the tip orientation of $^{249}$Bk,
ranging over a much wider energy domain, from $E_\mathrm{c.m.}=191$~MeV to $E_\mathrm{c.m.}=218$~MeV,
which is the highest energy we have computed.  
The tip orientation also leads to smaller maximum mass and charge transfer.
% compared to the side orientation of $^{249}$Bk.

\begin{figure}[!htb]
	\centerline{
		\includegraphics*[scale=0.3]{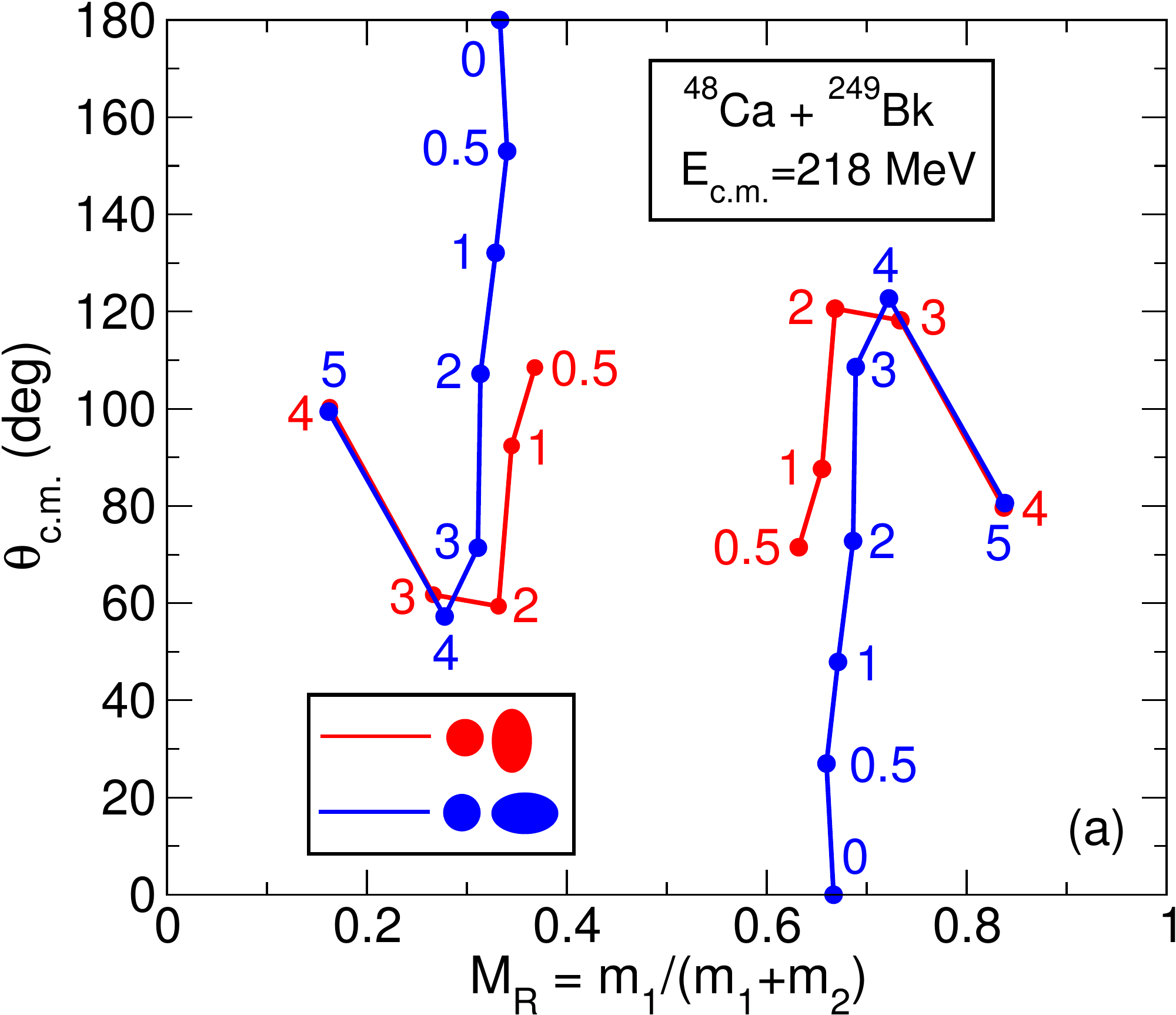}\hspace{0.2in}
		\includegraphics*[scale=0.3]{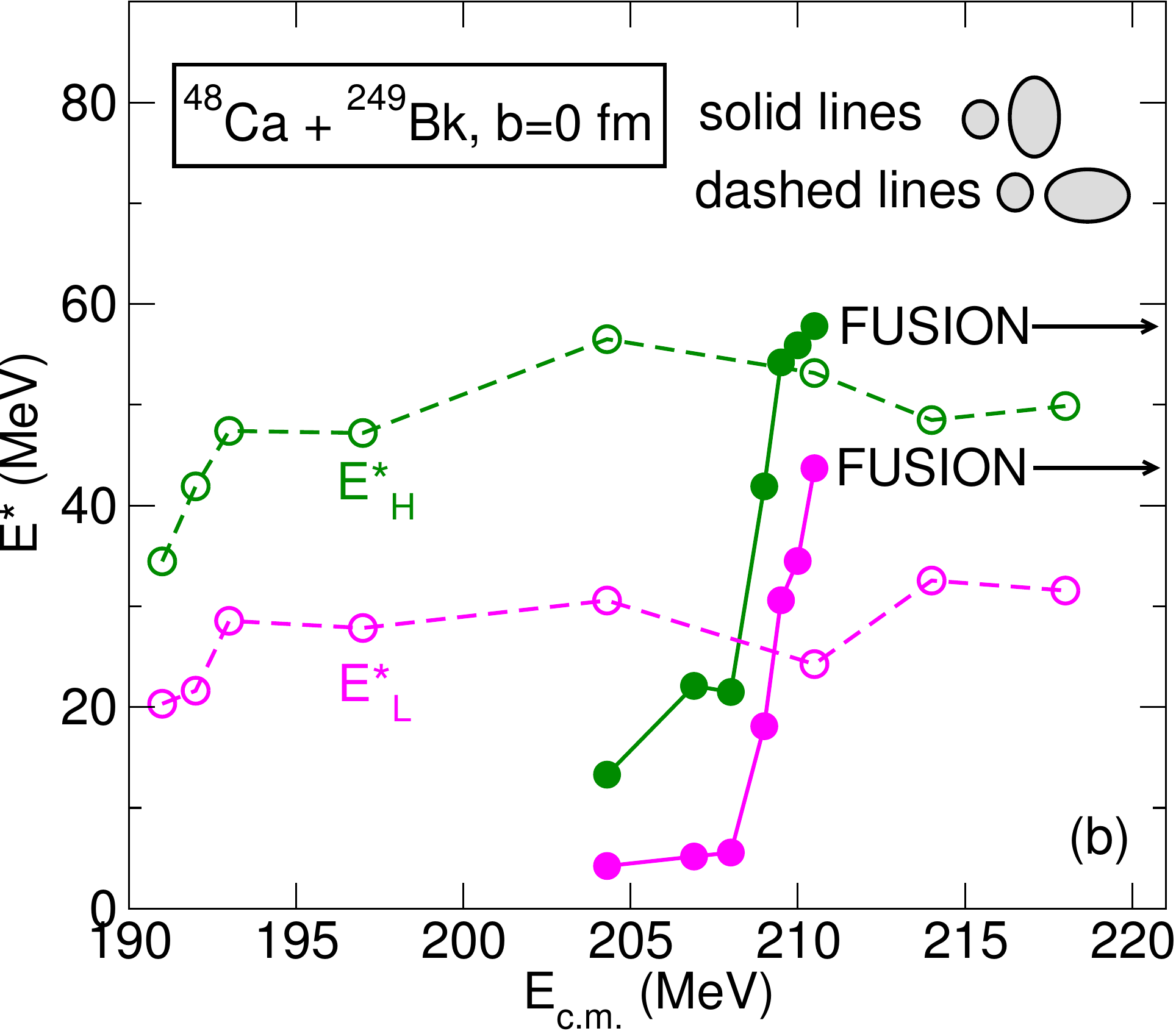}}
	\caption{\protect (a) Mass-angle distribution for the $^{48}$Ca+$^{249}$Bk system
		for the tip and side orientations of the $^{249}$Bk nucleus.
		(b) Excitation energy, $E^{*}$, of the heavy and light fragments
		as a function of $E_\mathrm{c.m.}$ for central collisions
		and for two orientations of the $^{249}$Bk nucleus.}
	\label{fig:fig1}
\end{figure}
In Fig.~\ref{fig:fig1}(a) we show the TDHF calculations of quasifission MADs for
$^{48}$Ca+$^{249}$Bk at $E_\mathrm{c.m.}=218$~MeV, corresponding to the two
orientations of the $^{249}$Bk nucleus.
The regions of MAD's near $M_R=0.2$ and $M_R=0.8$
correspond to elastic and quasielastic reactions, followed by transition to deep-inelastic
reactions and subsequently quasifission. The TDHF calculations predict 
where the transition from deep-inelastic to quasifission occurs,
 as well as the general behavior of the MADs. However, due to the fact
that TDHF is a deterministic theory, it will only give us the most probable outcome or path
for the  MADs rather than a full distribution.

Recently, we have developed an extension to TDHF theory via the use
of a density constraint to calculate the excitation energy of {\it each fragment} directly from the
TDHF density evolution.
This gives us new information on the repartition of the excitation energy between the heavy and light fragments
which is not available in standard TDHF calculations.
In Fig.~\ref{fig:fig1}b we show that the
heavy and light fragments  contain up to $60$~MeV and $45$~MeV (side
orientation) and up to $\sim55$~MeV and $\sim30$~MeV (tip orientation) of excitation energy,
respectively. %, for c.m. energies corresponding to quasifission.
It is interesting to note that the excitation energy follows qualitatively the mass and charge transfer behaviour observed in Fig.~\ref{fig:fig0}-b.

\begin{figure}[!htb]
	\centerline{
		\includegraphics*[scale=0.3]{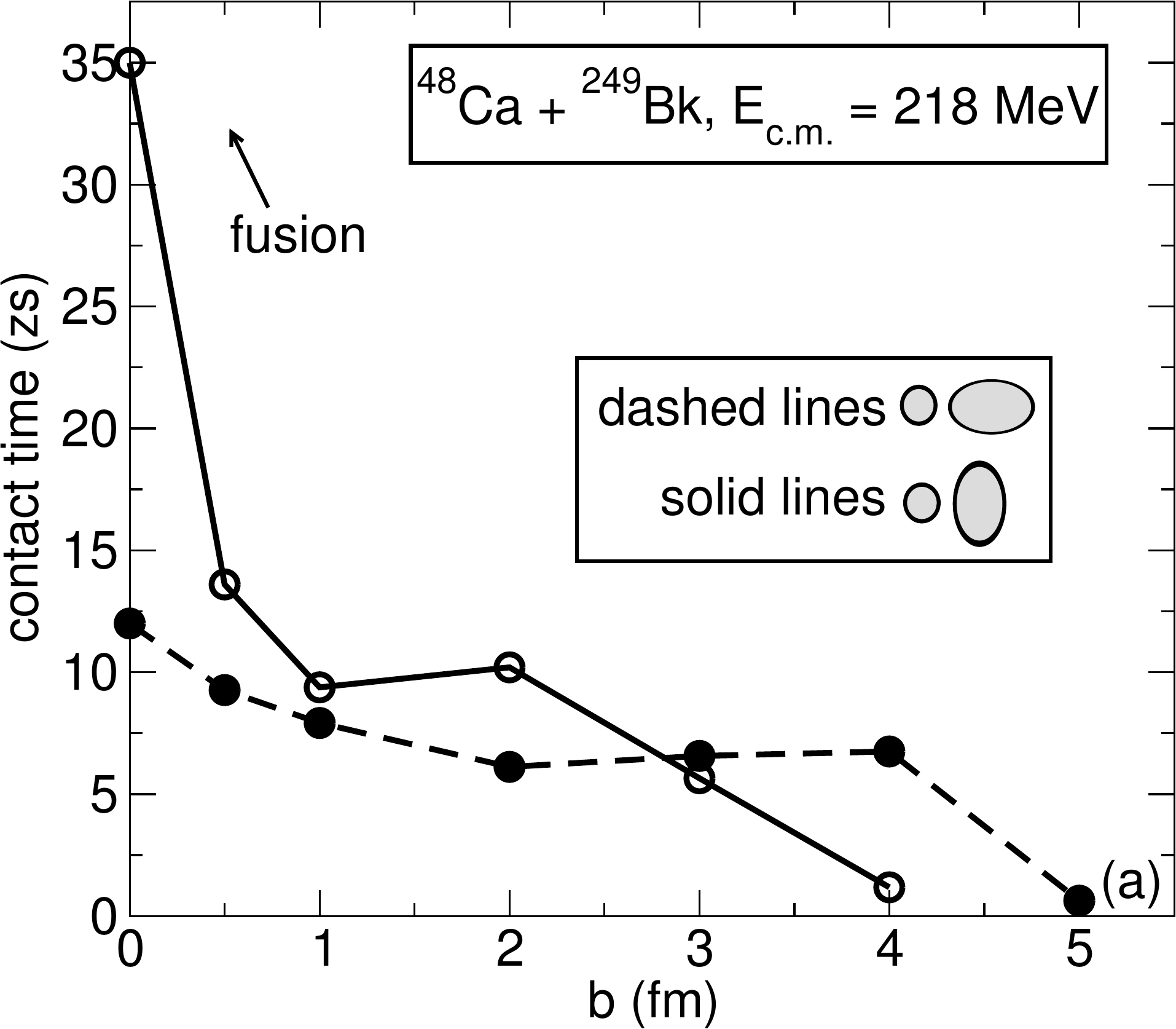}\hspace{0.2in}
		\includegraphics*[scale=0.3]{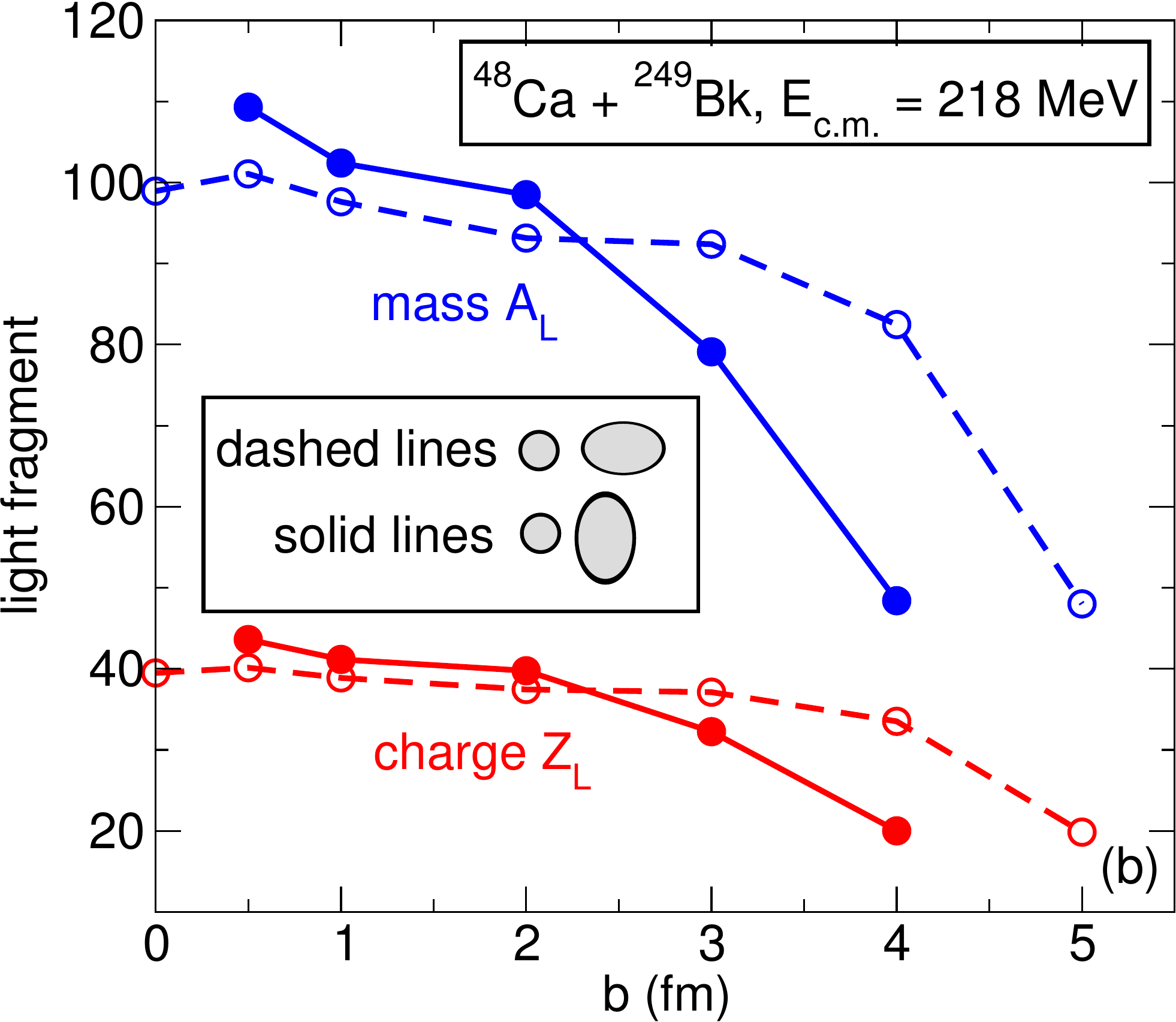}}
	\caption{\protect (a) Contact time and (b) mass and
		charge of the light fragment for the $^{48}$Ca+$^{249}$Bk system as a function of impact parameter for the two
		orientations of the $^{249}$Bk nucleus calculated at the c.m. energy of $E_{\mathrm{c.m.}}= 218$~MeV.}
	\label{fig:fig2}
\end{figure}

Figure~\ref{fig:fig2}a shows the contact time as a function of impact parameter at 
$E_{\mathrm{c.m.}}= 218$~MeV. We see that the contact times are $8-30$~zs for impact parameters
resulting in quasifission and falls sharply for fragments produced in deep-inelastic
collisions. 
These contact times are in agreement with typical quasi-fission times obtained 
from interpretations of experimental mass-angle distributions of the fragments \cite{Toke1985,Rietz2011,Rietz2013}. 

Figure~\ref{fig:fig2}b show the mass and charge of the light fragment for the $^{48}$Ca+$^{249}$Bk system as a function of impact parameter for the two
orientations of the $^{249}$Bk nucleus calculated at $E_{\mathrm{c.m.}}= 218$~MeV. As expected quasifission is identified with large mass and charge transfer,
in this case corresponding to the doubling of the charge from 20 to 40 and mass from 48
to 100.
It is also interesting to note the slightly atypical value of the contact time at impact parameter
$b=2$~fm in Fig.~\ref{fig:fig2}a in comparison to the neighboring impact parameters.
Fig.~\ref{fig:fig2}b shows that in this region
the light fragment is a neutron rich Zr isotope with $A\approx 102-106$.
The microscopic evolution of the shell structure seems to have a tendency to
form a composite with a longer lifetime when the light fragment is in this region.
Similar observations were made in $^{40,48}$Ca$+^{238}$U quasifission study~\cite{oberacker2014}. 
A possible explanation is the presence of strongly bound deformed isotopes of Zr in this
region~\cite{oberacker2003,blazkiewicz2005}.

\section{Summary}
We have shown recent quasifission results for the $^{48}$Ca+$^{249}$Bk system. Further
calculations are underway to obtain a full range of observables including mass-angle
distributions and fragment TKE's.
Recent TDHF calculations of phenomena related to superheavy element searches show that TDHF can be
a valuable tool for elucidating some of the underlying physics for these reactions.
As a microscopic theory with no free parameters, where the effective nucleon-nucleon
interaction is only fitted to the static properties of a few nuclei, these results
are very promising.

% Acknowledgement
\section{Acknowledgments}
This work has been supported by the U.S. Department of Energy under grant No.
DE-SC0013847 with Vanderbilt University and by the
Australian Research Council Grant No. FT120100760.

% References
%\bibliographystyle{myamsplain}
%\bibliography{AIP_proc}

\end{document}